\documentclass[a4paper]{article}
\usepackage{graphicx}
\usepackage{amsmath}
\usepackage{amssymb}

\begin{document}

\newcommand{\bin}[2]{\left(\begin{array}{c} \!\!#1\!\!\! \\ \!\! #2 \!\!\! \end{array}\right)}
\newcommand{\der}[2]{\frac{\partial #1}{\partial #2}}
\newcommand{\calE}{\mathcal{E}}
\newcommand{\h}{\mathrm{H}}
\newcommand{\he}{\mathrm{He}}
\newcommand{\threejm}[6]{\left(\begin{array}{ccc}#1 & #2 & #3 \\ #4 & #5 & #6 \end{array}\right)}

%===============================================================================
%	Detailed computation of hot-plasma atomic spectra  
%===============================================================================

\huge

\vspace{3cm}

\begin{center}
Detailed computation of hot-plasma atomic spectra
\end{center}

\vspace{0.5cm}

\large

\begin{center}
Jean-Christophe Pain$^{a,}$\footnote{jean-christophe.pain@cea.fr (corresponding author)}, Franck Gilleron$^a$ and Thomas Blenski$^b$
\end{center}

\vspace{0.2cm}

\normalsize

\begin{center}
$^a$CEA, DAM, DIF, F-91297 Arpajon, France

$^b$CEA, DSM, IRAMIS, F-91191 Gif-sur-Yvette, France
\end{center}

\vspace{0.5cm}

%===============================================================================
%		ABSTRACT
%===============================================================================

\begin{center}
{\bf Abstract}
\end{center}

We present recent evolutions of the detailed opacity code SCO-RCG which combines statistical modelings of levels and lines with fine-structure calculations. The code now includes the Partially Resolved Transition Array model, which allows one to replace a complex transition array by a small-scale detailed calculation preserving energy and variance of the genuine transition array and yielding improved high-order moments. An approximate method for studying the impact of strong magnetic field on opacity and emissivity was also recently implemented. The Zeeman line profile is modeled by fourth-order Gram-Charlier expansion series, which is a Gaussian multiplied by a linear combination of Hermite polynomials. Electron collisional line broadening is often modeled by a Lorentzian function and one has to calculate the convolution of a Lorentzian with Gram-Charlier distribution for a huge number of spectral lines. Since the numerical cost of the direct convolution would be prohibitive, we propose, in order to obtain the resulting profile, a fast and precise algorithm, relying on a representation of the Gaussian by cubic splines.

\begin{center}
{\bf PACS}
\end{center}

\begin{center}
32.30.-r, 32.70.-n, 32.60.+i
\end{center}

\begin{center}
{\bf Keywords}
\end{center}

\begin{center}
Hot plasmas - Radiative properties - Opacity - Fine structure - Statistical methods - E1 lines
\end{center}

\clearpage

\section{Introduction}

When atomic spectral lines coalesce into broad unresolved patterns due to physical broadening mechanisms (Stark effect, auto-ionization, etc.), they can be handled by so-called statistical methods (Bauche et al., 1988). Global characteristics - average energy, variance, asymmetry and sharpness - of level-energy, absorption or emission spectra  can be useful for their analysis and the investigation of their regularities (Bauche \& Bauche-Arnoult, 1987 \& 1990). Systematic studies of these average characteristics for transition arrays and applications to the interpretation of experimental spectra of high-temperature plasmas were initiated in (Moszkowski 1962, Bauche et al., 1979). The elaboration of the general group-diagrammatic summation method (Ginocchio, 1973) and its realization in computer codes (Kucas \& Karazija, 1993 \& 1995, Kucas et al., 2005, Karazija \& Kucas, 2013) opened up new possibilities for the use of global properties in atomic spectroscopy. On the other hand, some transition arrays exhibit a small number of lines that must be taken into account individually. Those lines are important for plasma diagnostics, interpretation of spectroscopy experiments and for calculating the Rosseland mean $\kappa_R$, important for radiation transport, and defined as

\begin{equation}
\frac{1}{\kappa_R}=\int_0^{\infty}\frac{1}{\kappa(h\nu)}\frac{\partial B_T(h\nu)}{\partial T}dh\nu\Bigg/\int_0^{\infty}\frac{\partial B_T(h\nu)}{\partial T}dh\nu,
\end{equation}

\noindent $h\nu$ being the incident photon energy, $\kappa(h\nu)$ the opacity including stimulated emission, $T$ the temperature and $B_T(h\nu)$ Planck's distribution function. The Rosseland mean is very sensitive to the gaps between lines in the spectrum. 

These are the reasons why we developed the hybrid opacity code SCO-RCG (Porcherot et al., 2011), which  combines statistical methods and fine-structure calculations, assuming local thermodynamic equilibrium. The main features of the code are described in section \ref{sec1}, the extension to the hybrid approach of the Partially Resolved Transition Array (PRTA) model (Iglesias \& Sonnad, 2012), which enables one to replace many statistical transition arrays by small-scale DLA (Detailed Line Accounting) calculations, is presented in section \ref{sec2} and comparisons with experimental spectra are shown and discussed in section \ref{sec3}. In section \ref{sec4}, an approximate modeling of Zeeman effect is proposed together with a fast numerical algorithm for the convolution of a Lorentzian function with Gram-Charlier expansion series, based on a cubic-spline representation of the Gaussian. 

\section{\label{sec1} Description of the code and effect of detailed lines}

In order to decide, for each transition array, whether a detailed treatment of lines is necessary or not and to determine the validity of statistical methods, the SCO-RCG code uses criteria to quantify the porosity (localized absence of lines) of transition arrays. The main quantity involved in the decision process is the ratio between the individual line width and the average energy gap between two lines in a transition array. Data required for the calculation of lines (Slater, spin-orbit and dipolar integrals) are provided by SCO (Superconfiguration Code for Opacity) code (Blenski et al., 2000), which takes into account plasma screening and density effects on the wave-functions. Then, level energies and lines are calculated by an adapted routine (RCG) of Cowan's atomic-structure code (Cowan, 1981) performing the diagonalization of the Hamiltonian matrix. Transition arrays for which a DLA treatment is not required or impossible are described statistically, by UTA (Unresolved Transition Array, Bauche-Arnoult et al., 1979), SOSA (Spin-Orbit Split Array, Bauche-Arnoult et al., 1985) or STA (Super Transition Array, Bar-Shalom et al., 1989) formalisms used in SCO. In SCO-RCG, the orbitals are treated individually up to a certain limit, consistent with Inglis-Teller limit (Inglis \& Teller, 1939), beyond which they are gathered in a single super-shell. The grouped orbitals are chosen so that they weakly interact with inner orbitals (this is why we sometimes name that super-shell ``Rydberg'' super-shell). The total opacity is the sum of photo-ionization, inverse Bremsstrahlung and Thomson scattering spectra calculated by SCO code and a photo-excitation spectrum in the form

\begin{equation}
\kappa\left(h\nu\right)=\frac{1}{4\pi\epsilon_0}\frac{\mathcal{N}}{A}\frac{\pi e^2h}{mc}\sum_{X\rightarrow X'}f_{X\rightarrow X'}\mathcal{P}_X\Psi_{X\rightarrow X'}(h\nu),
\end{equation}

\noindent where $h$ is Planck's constant, $\mathcal{N}$ the Avogadro number, $\epsilon_0$ the vacuum polarizability, $m$ the electron mass, $A$ the atomic number and $c$ the speed of light. $\mathcal{P}$ is a probability, $f$ an oscillator strength, $\Psi(h\nu)$ a profile and the sum $X\rightarrow X'$ runs over lines, UTA, SOSA or STAs of all ion charge states present in the plasma. Special care is taken to calculate appropriately the probability of $X$ (which can be either a level $\alpha J$, a configuration $C$ or a superconfiguration $S$) because it can be the starting point for different transitions (DLA, UTA, SOSA, STA). In order to ensure the normalization of probabilities, we introduce three disjoint ensembles: $\mathcal{D}$ (detailed levels $\alpha J$), $\mathcal{C}$ (configurations $C$ too complex to be detailed) and $\mathcal{S}$ (superconfigurations $S$ that do not reduce to ordinary configurations). The total partition function then reads

\begin{equation}
U_{\mathrm{tot}}=U\left(\mathcal{D}\right)+U\left(\mathcal{C}\right)+U\left(\mathcal{S}\right)\;\;\;\;\mathrm{with}\;\;\;\;\mathcal{D}\cap\mathcal{C}\cap\mathcal{S}=\emptyset,
\end{equation}

\noindent where each term is a trace over quantum states of the form Tr$\left[e^{-\beta\left(\hat{H}-\mu\hat{N}\right)}\right]$, where $\hat{H}$ is the Hamiltonian, $\hat{N}$ the number operator, $\mu$ the chemical potential and $\beta=1/\left(k_BT\right)$. The probabilities of the different species of the $N$-electron ion are

\begin{equation}\label{probs}
\mathcal{P}_{\alpha J}=\frac{1}{U_{\mathrm{tot}}}\left(2J+1\right)e^{-\beta\left(E_{\alpha J}-\mu N\right)},
\end{equation}

\noindent for a level belonging to $\mathcal{D}$,

\begin{equation}
\mathcal{P}_C=\frac{1}{U_{\mathrm{tot}}}\sum_{\alpha J\in C}\left(2J+1\right)e^{-\beta\left(E_{\alpha J}-\mu N\right)},
\end{equation}

\noindent for a configuration that can be detailed, 

\begin{equation}
\mathcal{P}_C=\frac{1}{U_{\mathrm{tot}}}g_C~e^{-\beta\left(E_C-\mu N\right)}
\end{equation}

\noindent for a configuration that can not be detailed (\emph{i.e.} belonging to $\mathcal{C}$) and

\begin{equation}
\mathcal{P}_S=\frac{1}{U_{\mathrm{tot}}}\sum_{C\in S}g_C~e^{-\beta\left(E_C-\mu N\right)}
\end{equation}

\noindent for a superconfiguration.

\begin{figure}
\vspace{10mm}
\begin{center}
\includegraphics[width=10cm]{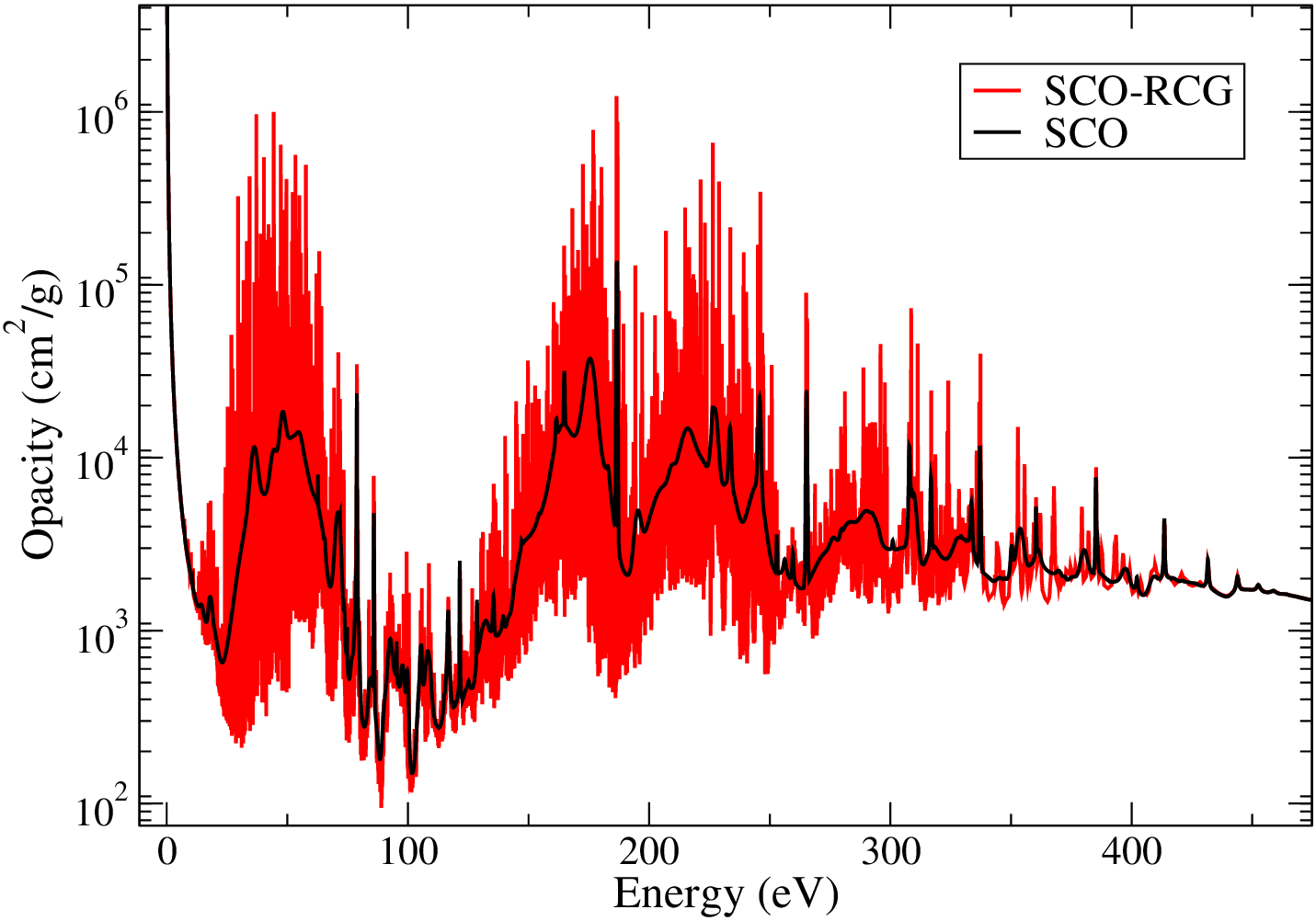}
\end{center}
\vspace{5mm}
\caption{(Color online) Comparison between the SCO-RCG and full-statistical (SCO) calculations for an iron plasma at $T$=50 eV and $\rho$=10$^{-3}$ g/cm$^2$. The Rosseland mean is equal to 1691 cm$^2$/g for the SCO calculation and to 1261 cm$^2$/g for SCO-RCG.\label{DLA_dilue}}
\end{figure}

We can see in Fig. \ref{DLA_dilue} that fine-structure calculations can have a strong impact on the Rosseland mean. The physical broadening mechanisms are the same for both calculations (statistical: SCO and detailed: SCO-RCG). The modelling of the (impact) collisional broadening relies on the Baranger formulation (Baranger, 1958) and expressions provided by Dimitrijevic and Konjevic (Dimitrijevic \& Konjevic, 1987) corrected by inelastic Gaunt factors similar, for high energies, to the ones proposed by Griem (1962 \& 1968). Ionic Stark effect is treated in the quasi-static approximation following an approach proposed by Rozsnyai (1977), corrected in order to reproduce the exact second-order moment of the electric micro-field distribution in the framework of the OCP (One Component Plasma) model (Iglesias et al., 1983). 

\begin{figure}
\vspace{10mm}
\begin{center}
\includegraphics[width=10cm]{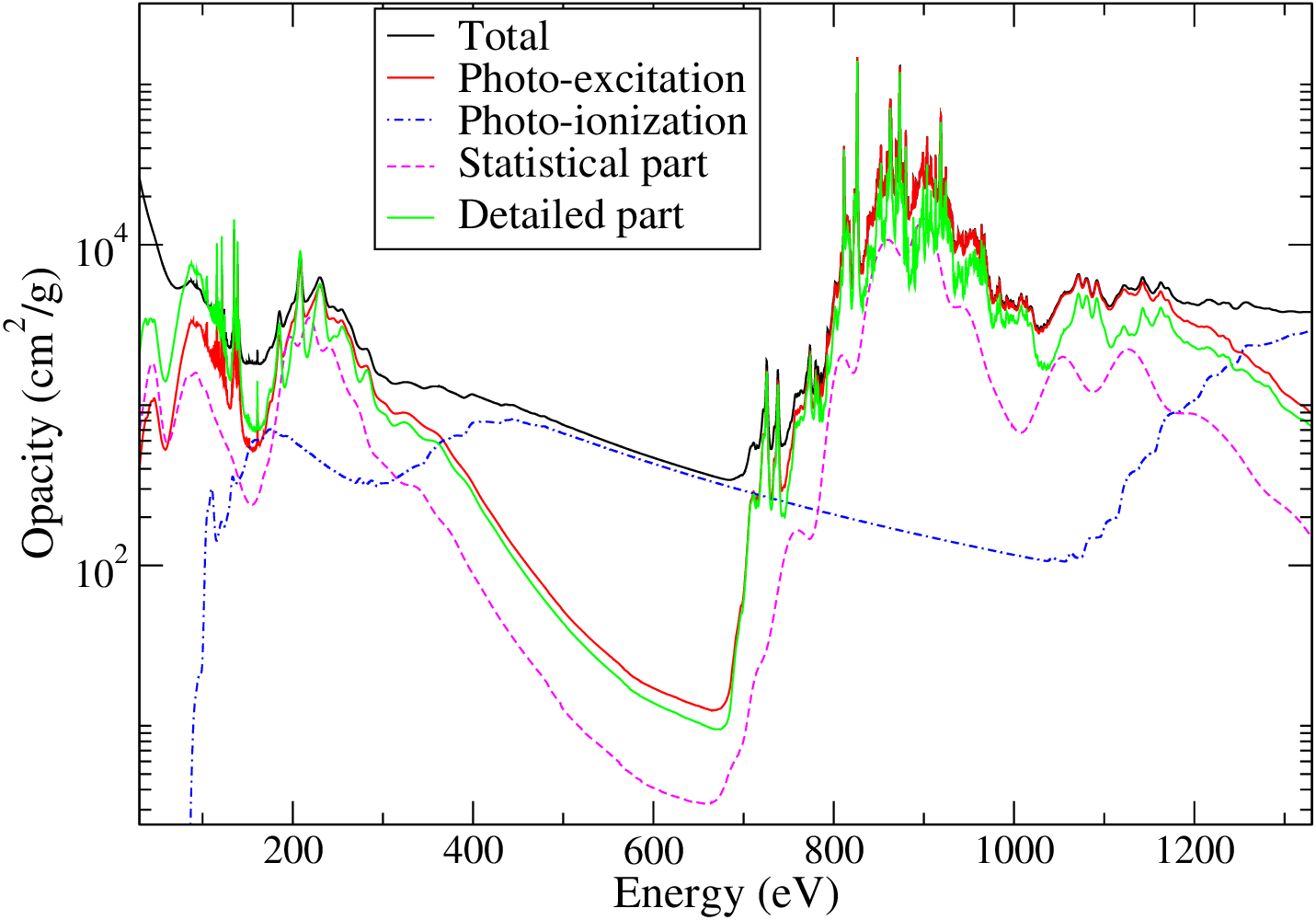}
\end{center}
\vspace{5mm}
\caption{(Color online) Different contributions to opacity calculated by SCO-RCG code for an iron plasma at $T$=193 eV and $\rho$=0.58 g.cm$^{-3}$ (boundary of the convective zone of the Sun). The maximum number of lines potentially detailed per transition array is chosen equal to 100.\label{Fe_BCZ_100_cut}}
\end{figure}

\begin{figure}
\vspace{5mm}
\begin{center}
\includegraphics[width=10cm]{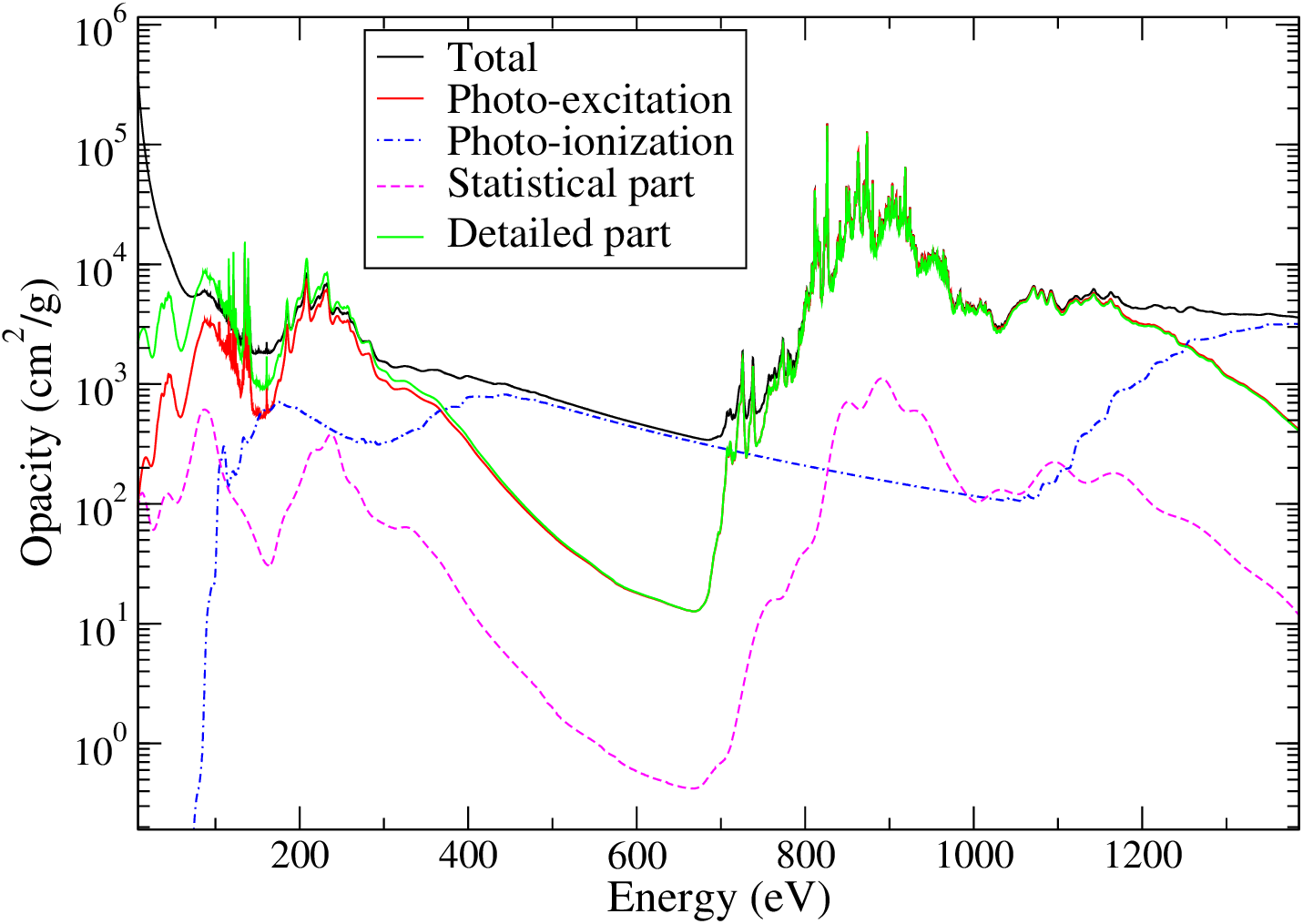}
\end{center}
\vspace{5mm}
\caption{(Color online) Different contributions to opacity calculated by SCO-RCG code for an iron plasma at $T$=193 eV and $\rho$=0.58 g.cm$^{-3}$ (boundary of the convective zone of the Sun). The maximum number of lines potentially detailed per transition array is chosen equal to 800000.\label{Fe_BCZ_800000_cut}}
\end{figure}

\begin{figure}
\vspace{10mm}
\begin{center}
\includegraphics[width=10cm]{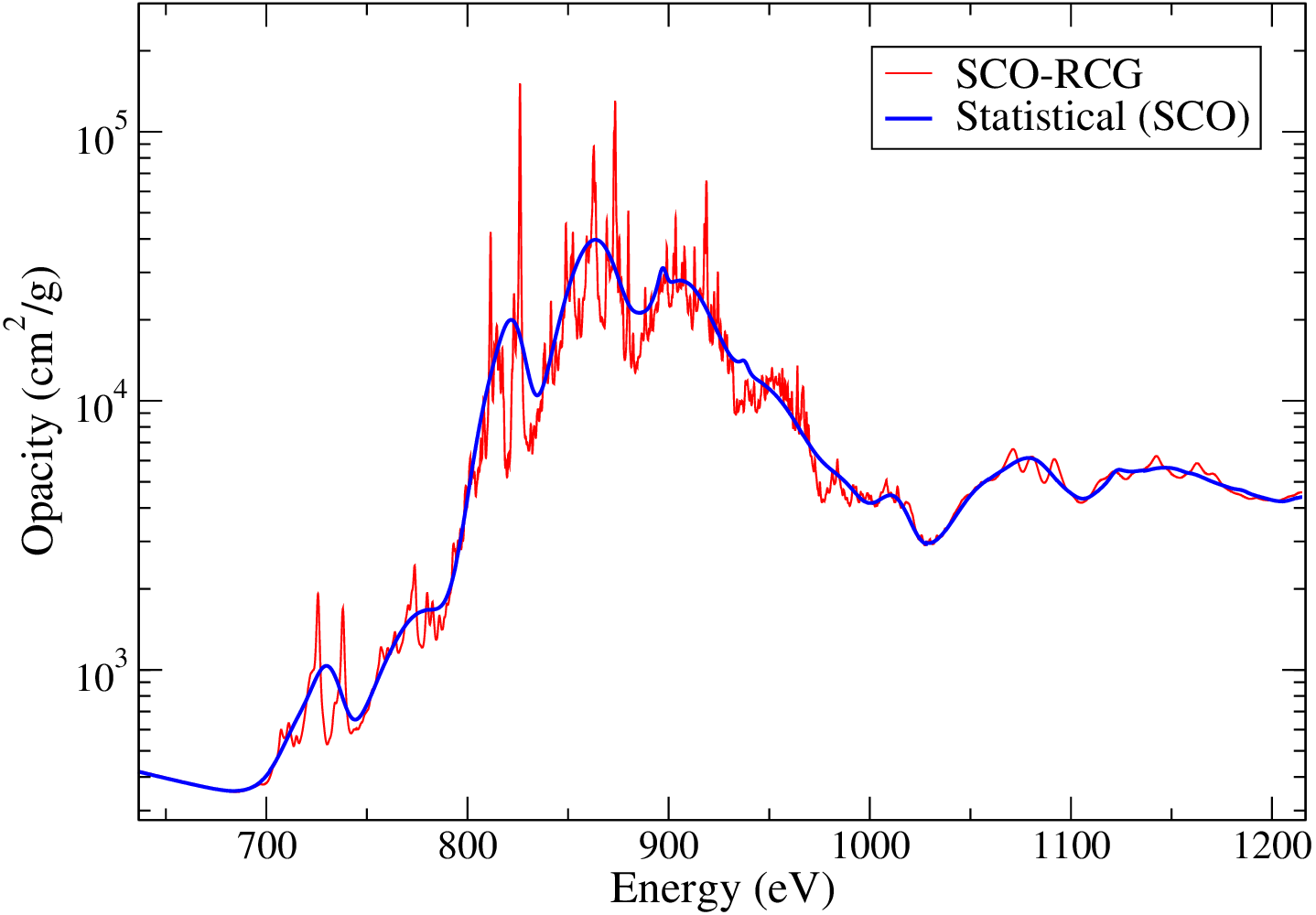}
\end{center}
\vspace{5mm}
\caption{(Color online) Comparison between the full-statistical (SCO) spectrum and SCO-RCG around the maximum of the opacity bump in the conditions of Fig. \ref{Fe_BCZ_100_cut} (boundary of the convective zone of the Sun). The maximum number of lines potentially detailed per transition array is chosen equal to 800000.\label{Fe_BCZ_hybrid+sco_cut}}
\end{figure}

The code can be useful for astrophysical applications (Gilles et al., 2011, Turck-Chi\`eze et al., 2011). Figures \ref{Fe_BCZ_100_cut}, \ref{Fe_BCZ_800000_cut} and \ref{Fe_BCZ_hybrid+sco_cut} represent the different contributions to opacity (DLA, statistical and PRTA) for an iron plasma in conditions corresponding to the boundary of the convective zone of the Sun, with a maximum imposed number $N_{\mathrm{max}}$ of detailed lines per transition array equal to 100 and 800000 respectively. As expected, when $N_{\mathrm{max}}$ increases, the statistical part becomes smaller. In Fig. \ref{Fe_BCZ_100_cut}, the detailed part is obviously not sufficiently larger than the statistical part around $h\nu$=850 eV.

\begin{figure}
\vspace{10mm}
\begin{center}
\includegraphics[width=10cm]{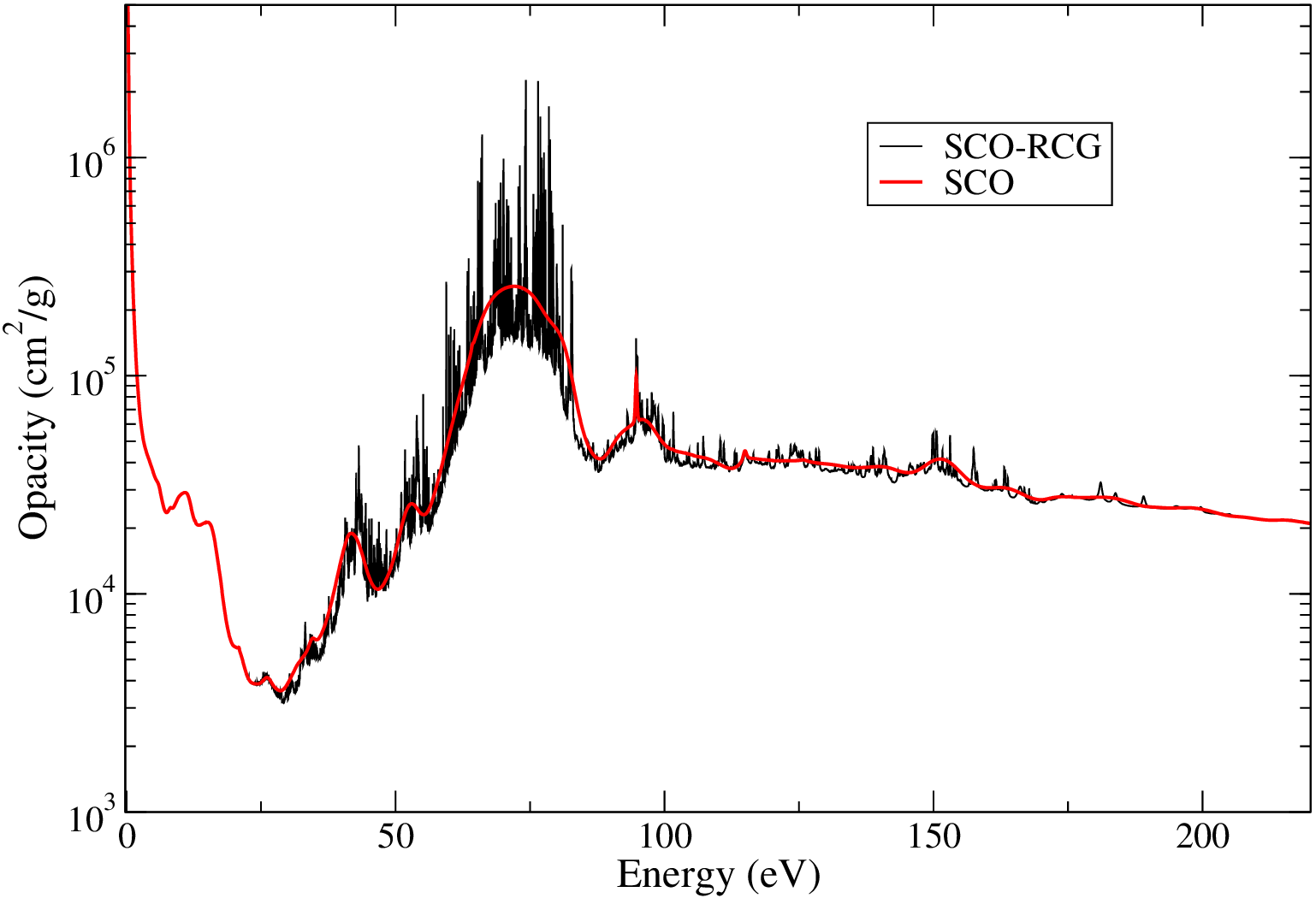}
\end{center}
\vspace{5mm}
\caption{(Color online) Iron opacity at $T$=15 eV and $\rho$= 4 10$^{-3}$ g.cm$^{-3}$. Comparison between the full-statistical (SCO) and SCO-RCG calculations. The maximum number of lines potentially detailed per transition array is chosen equal to 800000.\label{fer_gilles_15}}
\end{figure}

\begin{figure}
\vspace{10mm}
\begin{center}
\includegraphics[width=10cm]{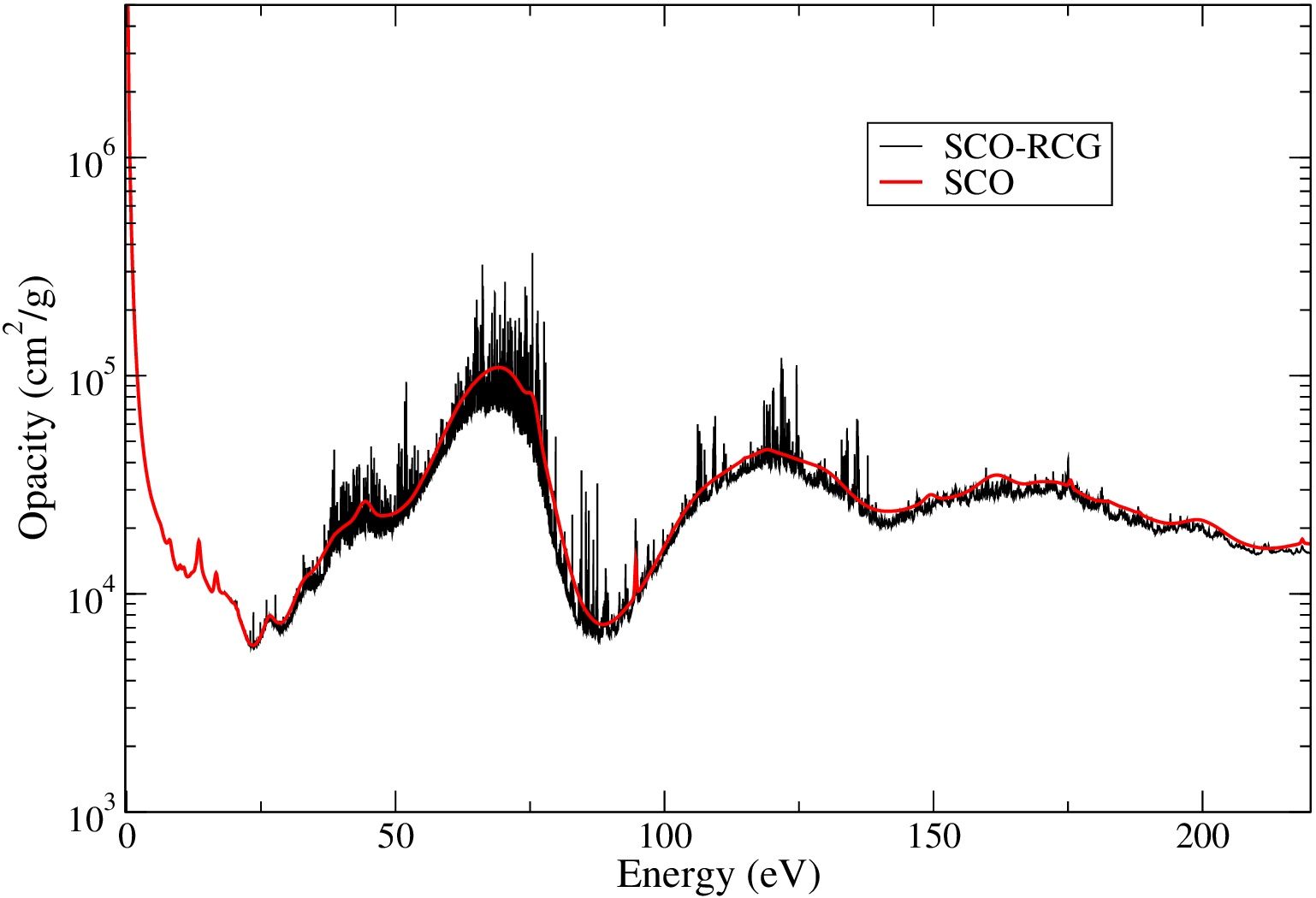}
\end{center}
\vspace{5mm}
\caption{(Color online) Iron opacity at $T$=27 eV and $\rho$= 4 10$^{-3}$ g.cm$^{-3}$. Comparison between the full-statistical (SCO) and SCO-RCG calculations. The maximum number of lines potentially detailed per transition array is chosen equal to 800000.\label{fer_gilles_27}}
\end{figure}

\begin{figure}
\vspace{10mm}
\begin{center}
\includegraphics[width=10cm]{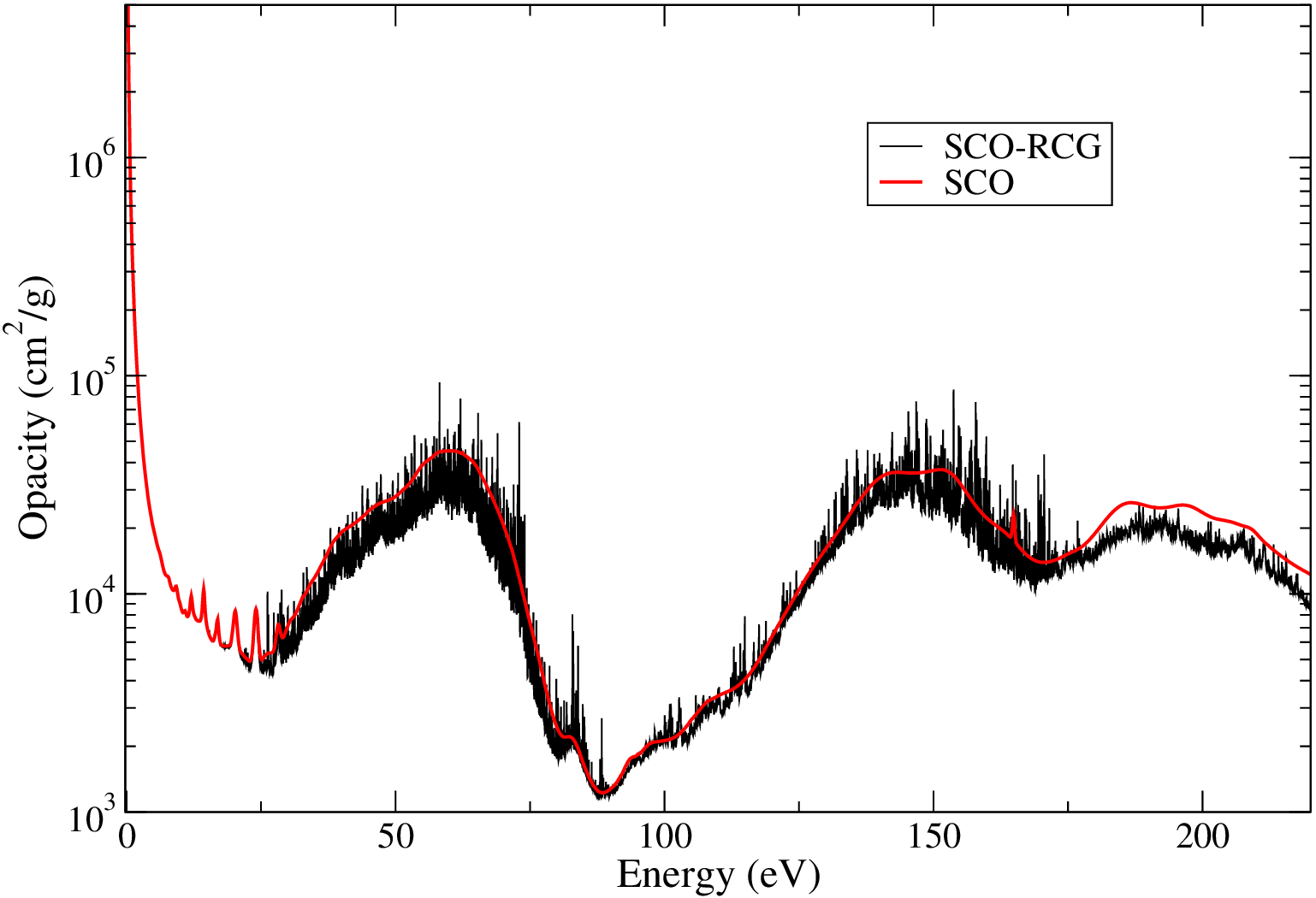}
\end{center}
\vspace{5mm}
\caption{(Color online) Iron opacity at $T$=38 eV and $\rho$= 4 10$^{-3}$ g.cm$^{-3}$. Comparison between the full-statistical (SCO) and SCO-RCG calculations. The maximum number of lines potentially detailed per transition array is chosen equal to 800000.\label{fer_gilles_38}}
\end{figure}

As shown on figures \ref{fer_gilles_15}, \ref{fer_gilles_27} and \ref{fer_gilles_38}, the statistical calculation (SCO) may depart significantly from the detailed one (SCO-RCG), and the differences are essentially the signature of the porosity of transition arrays. The density being quite low ($\rho$= 4 10$^{-3}$ g.cm$^{-3}$), the lines emerge clearly in the spectrum. These conditions are accessible to laser spectroscopy experiments (see Sec. \ref{sec3}) and we can see that the opacity changes notably with temperature. Therefore, even if the quantity which is measured in absorption point-projection-spectroscopy experiments is not the opacity itself, but the transmission (see Sec. \ref{sec3}), one may expect to have a reliable idea of the plasma temperature during the measurement.

\begin{figure}
\vspace{5mm}
\begin{center}
\includegraphics[width=10cm]{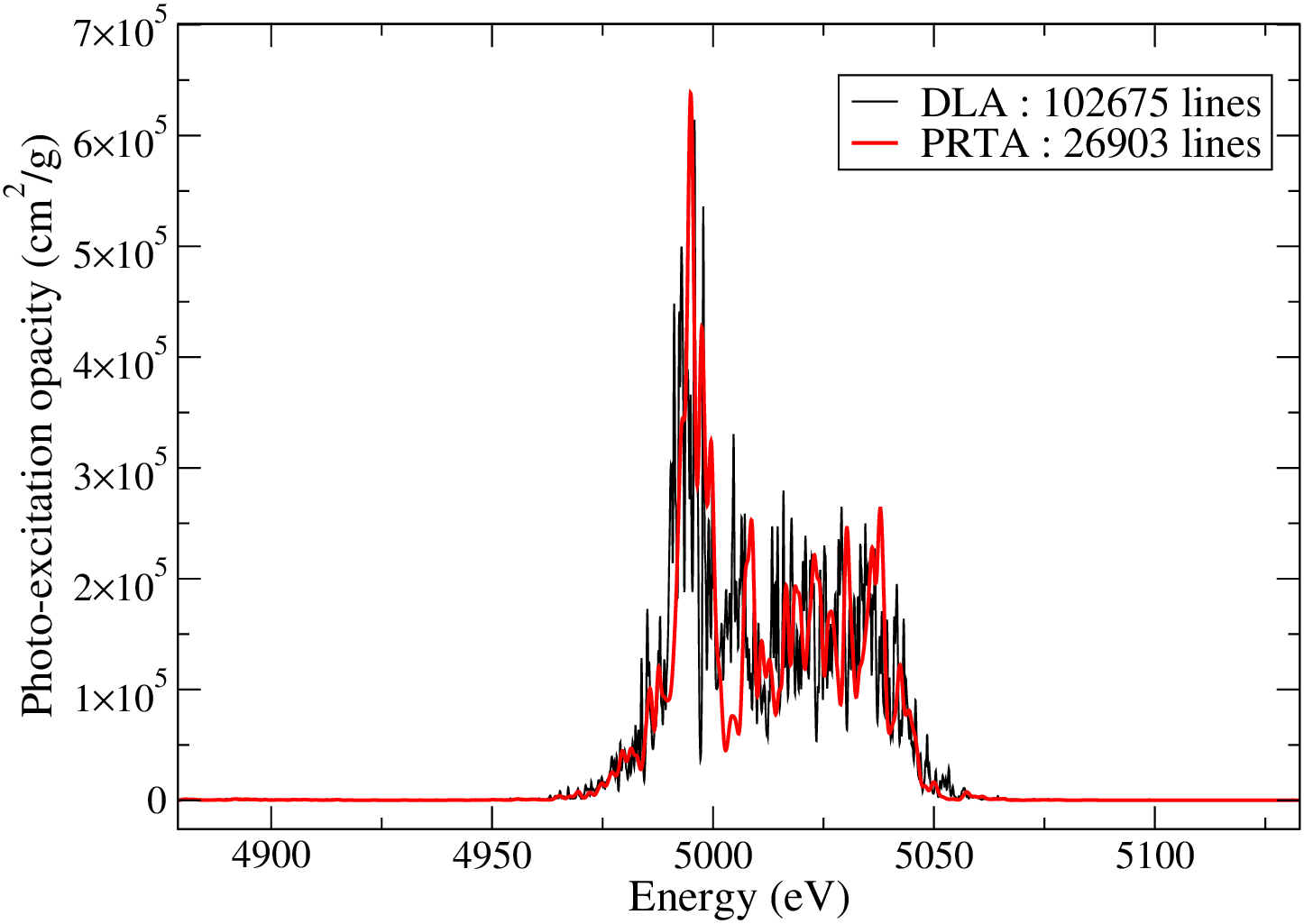}
\end{center}
\vspace{5mm}
\caption{(Color online) Comparison between two SCO-RCG calculations relying respectively on DLA and PRTA treatments of lines for transition arrays $3p_{3/2}\rightarrow 5s$ in a Hg plasma at $T$=600 eV and $\rho$=0.01 g/cm$^3$. The DLA calculation contains 102 675 lines and the PRTA 26 903 lines.\label{Hg_prta3.1_bis}}
\end{figure}

\begin{figure}
\vspace{5mm}
\begin{center}
\includegraphics[width=10cm]{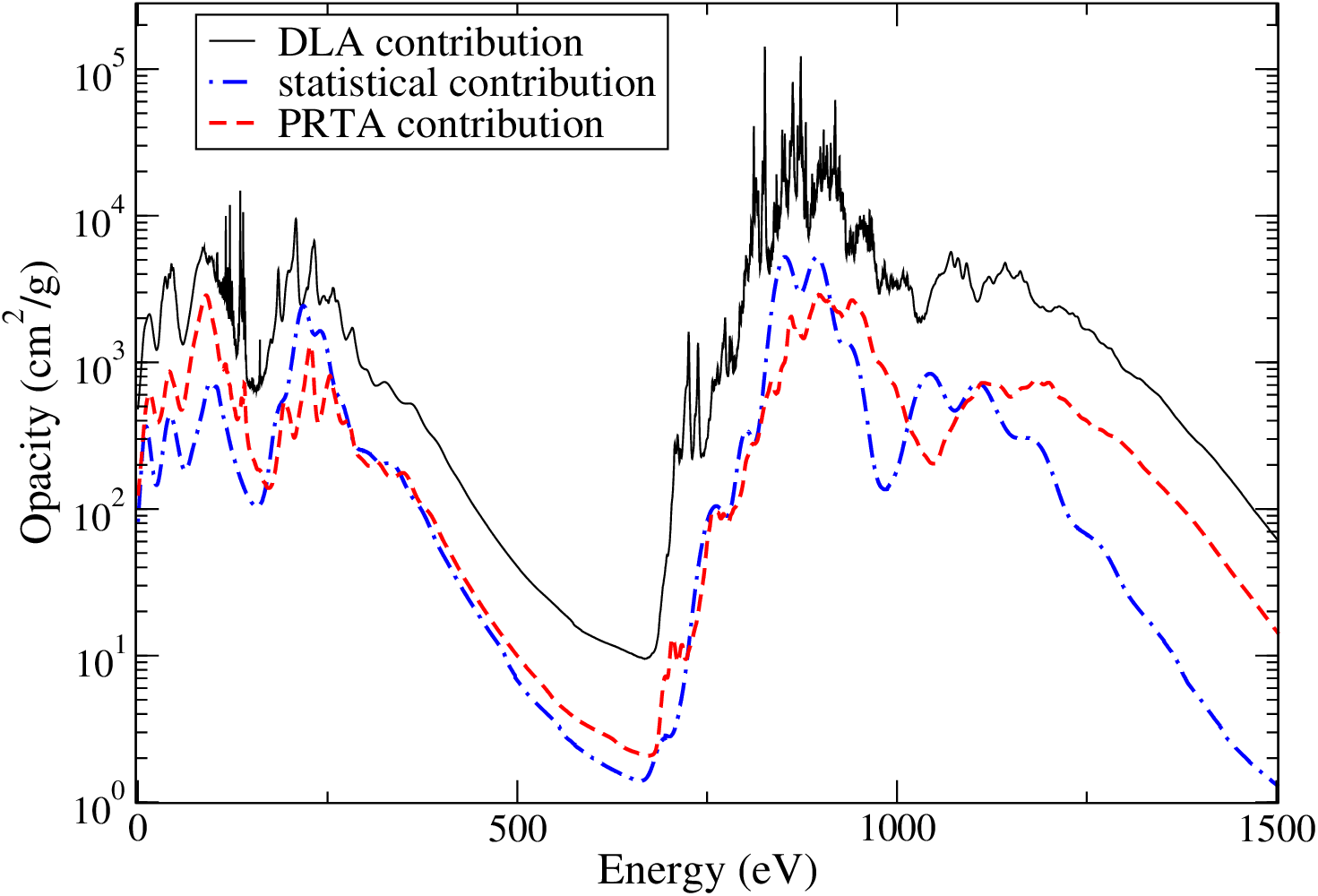}
\end{center}
\vspace{5mm}
\caption{(Color online) The three independent contributions to photo-excitation calculated by SCO-RCG code for an iron plasma at $T$=193 eV and $\rho$=0.58 g.cm$^{-3}$ (boundary of the convective zone of the Sun).\label{figure2}}
\end{figure}

\begin{figure}
\vspace{5mm}
\begin{center}
\includegraphics[width=10cm]{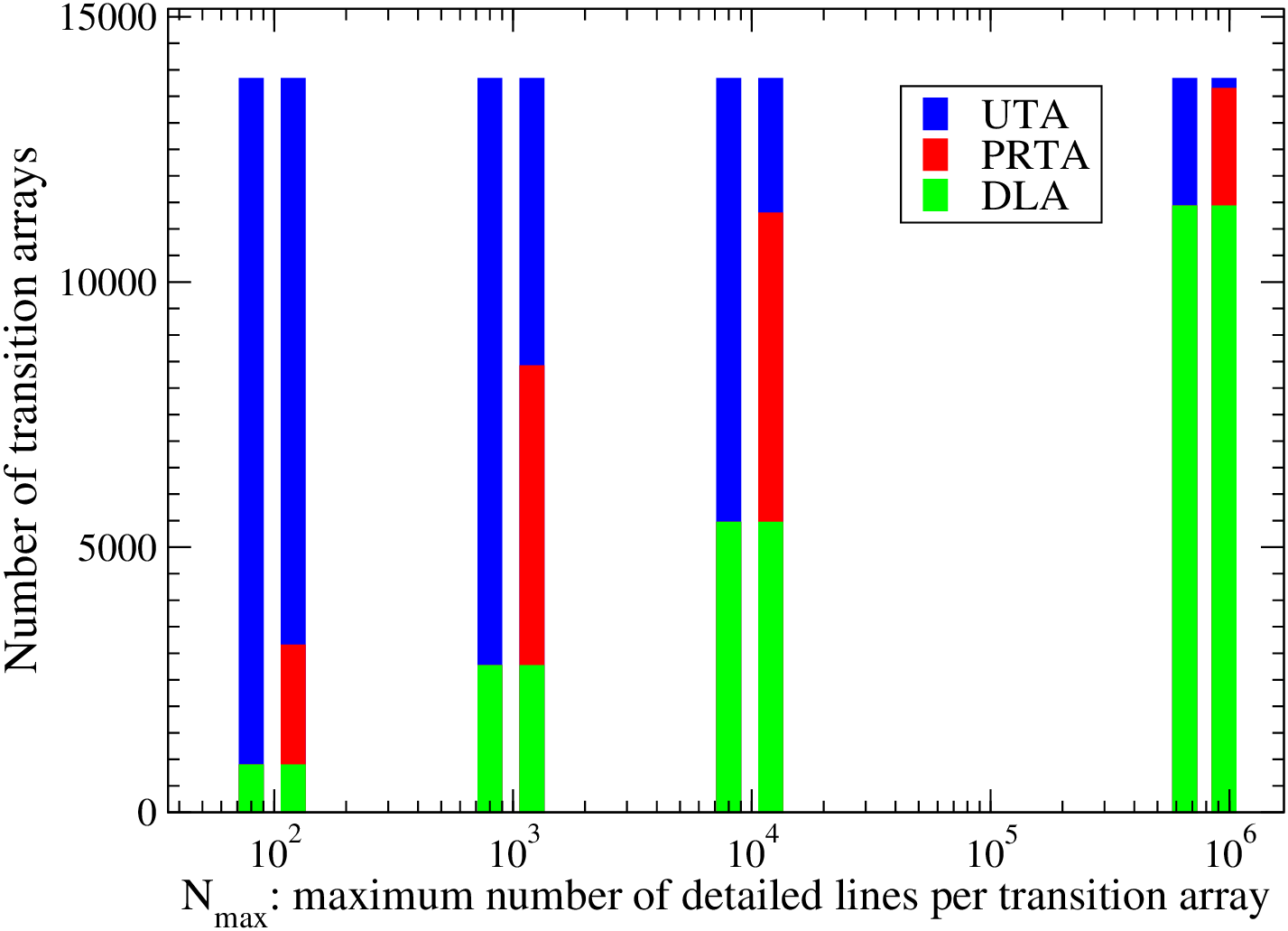}
\end{center}
\vspace{5mm}
\caption{(Color online) Number of detailed transition arrays (DLA), partially resolved transition arrays (PRTA) and unresolved transition arrays (UTA) for different values of the maximum number of detailed lines imposed : 10$^2$, 10$^3$, 10$^4$ and $10^6$. For each case, two histograms are displayed: in the first one, the detailed calculations are only pure DLA and in the second one they can be either DLA or PRTA.\label{dla-prta_2}}
\end{figure}

\section{\label{sec2} Adaptation of the PRTA model to the hybrid approach}

In order to complement DLA efforts, the code was recently improved (Pain et al., 2013a \& 2015) with the PRTA (Partially Resolved Transition Array) model (Iglesias \& Sonnad, 2012), which may replace the single feature of a UTA by a small-scale detailed transition array that conserves the known transition-array properties (energy and variance) and yields improved higher-order moments. In the PRTA approach, open subshells are split in two groups. The main group includes the active electrons and those electrons that couple strongly with the active ones. The other subshells are relegated to the secondary group. A small-scale DLA calculation is performed for the main group (assuming therefore that the subshells in the secondary group are closed) and a statistical approach for the secondary group assigns the missing UTA variance to the lines. In the case where the transition $C\rightarrow C'$ is a UTA that can be replaced by a PRTA (see Fig. \ref{Hg_prta3.1_bis}), its contribution to the opacity is modified according to

\begin{equation}
f_{C\rightarrow C'}~\mathcal{P}_C~\Psi_{C\rightarrow C'}(h\nu)\approx\sum_{\bar{\alpha}\bar{J}\rightarrow\bar{\alpha'}\bar{J'}}f_{\bar{\alpha}\bar{J}\rightarrow\bar{\alpha'}\bar{J'}}~\mathcal{P}_{\bar{\alpha}\bar{J}}~\Psi_{\bar{\alpha}\bar{J}\rightarrow\bar{\alpha'}\bar{J'}}(h\nu),
\end{equation}

\noindent where the sum runs over PRTA lines $\bar{\alpha}\bar{J}\rightarrow\bar{\alpha'}\bar{J'}$ between pseudo-levels of the reduced configurations, $f_{\bar{\alpha}\bar{J}\rightarrow\bar{\alpha'}\bar{J'}}$ is the corresponding oscillator strength and $\Psi_{\bar{\alpha}\bar{J}\rightarrow\bar{\alpha'}\bar{J'}}$ is the line profile augmented with the statistical width due to the other (non included) spectator subshells. The probability of the pseudo-level $\bar{\alpha}\bar{J}$ of configuration $\bar{C}$ reads

\begin{equation}
\mathcal{P}_{\bar{\alpha}\bar{J}}=\frac{\left(2\bar{J}+1\right)e^{-\beta\left(E_{\bar{\alpha}\bar{J}}-\mu N\right)}}{\sum_{\bar{\alpha}\bar{J}\in\bar{C}}\left(2\bar{J}+1\right)e^{-\beta \left(E_{\bar{\alpha}\bar{J}}-\mu N\right)}}\times\mathcal{P}_C
\end{equation}

\noindent with ensures that $\sum_{\bar{\alpha}\bar{J}\in\bar{C}}\mathcal{P}_{\bar{\alpha}\bar{J}}=\mathcal{P}_C$, where $\mathcal{P}_C$ is the probability of the genuine configuration given in Eq. (\ref{probs}).

Figure \ref{figure2} represents the different contributions to opacity (DLA, statistical and PRTA) for an iron plasma in conditions corresponding to the boundary of the convective zone of the Sun. We can see that the PRTA contribution is of the same order of magnitude here as the statistical one. The calculation was performed with a maximum imposed of 10 000 detailed lines per transition arrays. We can see in Fig. \ref{dla-prta_2} that for each value of the maximum number of lines that can be detailed ($N_{\mathrm{max}}$), some UTA are replaced by PRTA transition arrays. Of course, the number of remaining UTA decreases with $N_{\mathrm{max}}$.

\section{\label{sec3} Interpretation of experimental spectra}

\begin{figure}
\vspace{5mm}
\begin{center}
\includegraphics[width=10cm]{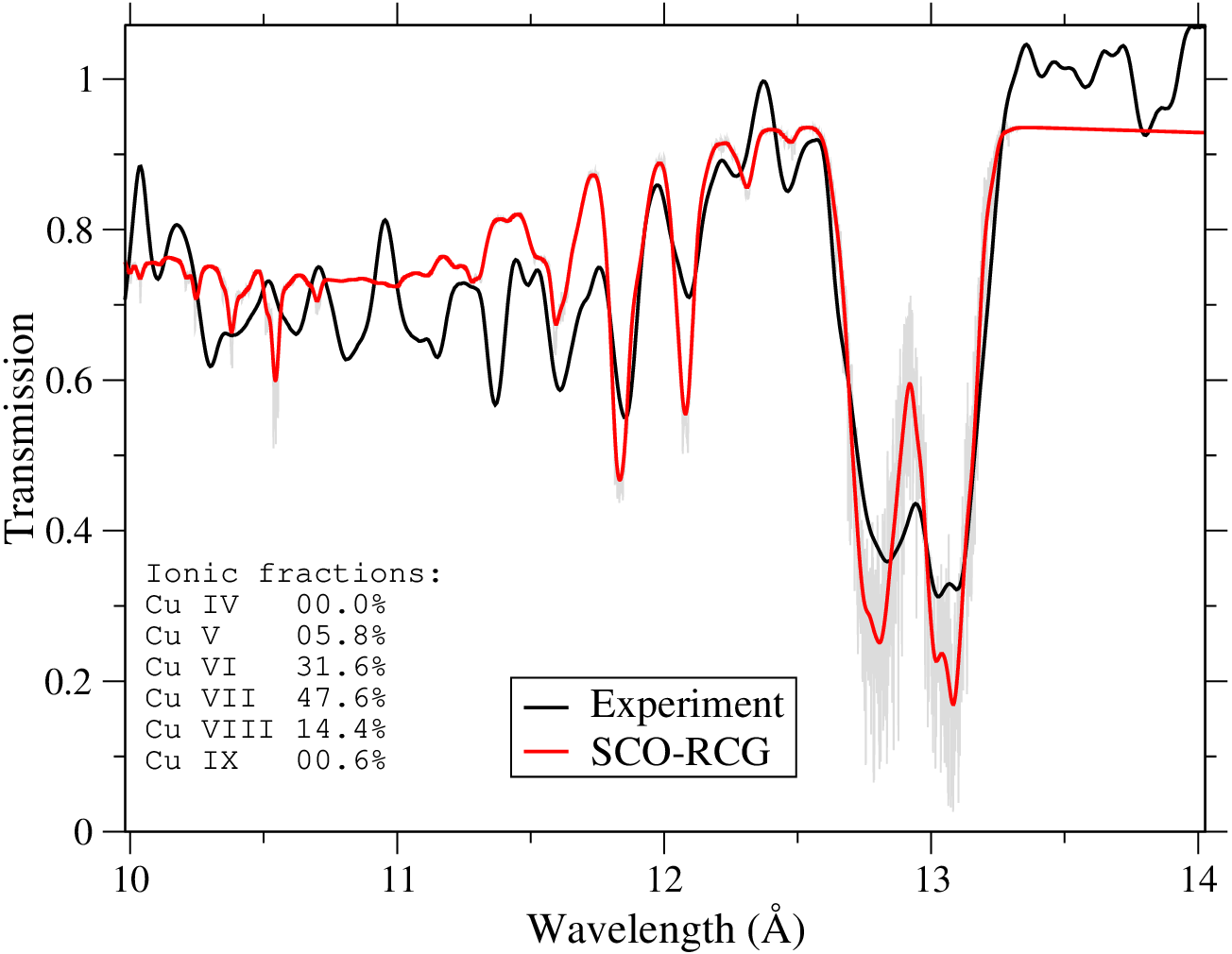}
\end{center}
\vspace{5mm}
\caption{(Color online) Interpretation with SCO-RCG code of the copper spectrum ($2p\rightarrow 3d$ transitions) measured by Loisel et al. (Loisel et al., 2009, Blenski et al., 2011a \& 2011b). The temperature is $T$=16 eV and the density $\rho$=5 10$^{-3}$ g.cm$^3$.\label{figure5}}
\end{figure}

\begin{figure}
\vspace{5mm}
\begin{center}
\includegraphics[width=14cm]{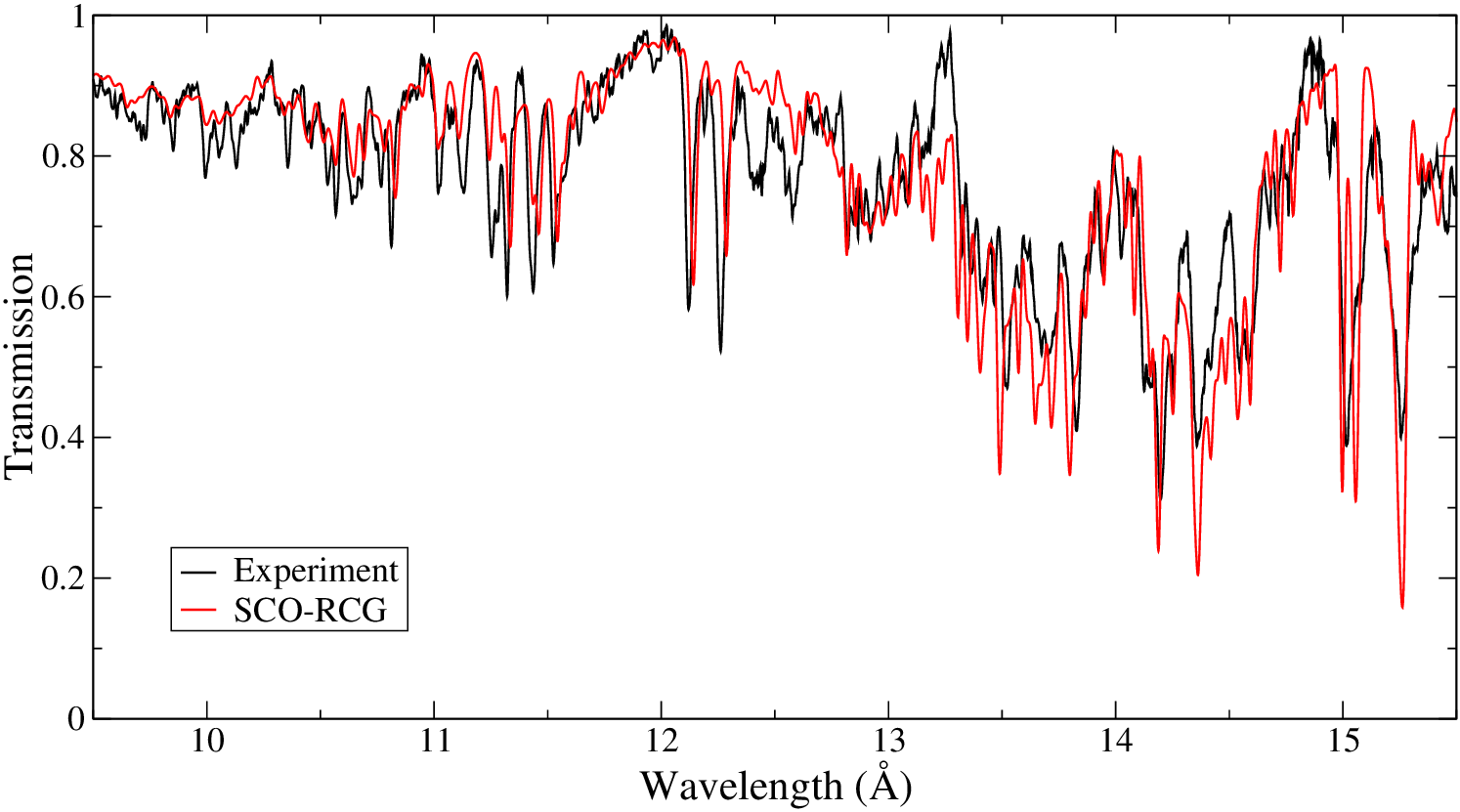}
\end{center}
\vspace{5mm}
\caption{(Color online) Interpretation with SCO-RCG code of the iron spectrum ($2p\rightarrow 3d$ transitions) measured by Bailey et al. (2009). The temperature is $T$=150 eV and the density $\rho$=0.058 g.cm$^3$.\label{figure4}}
\end{figure}

The SCO-RCG code has been successfully compared to several absorption and emission experimental spectra, measured in experiments at several laser (Fig. \ref{figure5}) or Z-pinch facilities (Fig. \ref{figure4}). The comparisons show the relevance of the hybrid model and the necessity to carry out detailed calculations instead of full statistical calculations. As mentioned in Sec. \ref{sec1}), the quantity which is measured experimentally is the transmission, related to the opacity by Beer-Lambert-Bouguer's law:

\begin{equation}\label{blb}
T(h\nu)=e^{-\rho L \kappa(h\nu)},
\end{equation}

\noindent where $L$ is the thickness of the sample. The relation (\ref{blb}) between transmission and opacity is valid under the assumption that the material is optically thin and that re-absorption processes are neglected. 

In SCO-RCG, configuration interaction is limited to electrostatic one between relativistic sub-configurations ($n\ell j$ orbitals) belonging to a non-relativistic configuration ($n\ell$ orbitals), namely ``relativistic configuration interaction''. That effect has a strong impact on the ratio of the two relativistic substructures of the $2p\rightarrow 3d$ transition on Fig. \ref{figure5}.

\clearpage

\section{\label{sec4} Statistical modeling of Zeeman effect}

\subsection{Determination of the moments}

\noindent Quantifying the impact of a magnetic field on spectral line shapes is important in astrophysics, in inertial confinement fusion (ICF) or for Z-pinch experiments. Because the line computation becomes even more tedious in that case, we propose, in order to avoid the diagonalization of the Zeeman Hamiltonian, to describe Zeeman patterns in a statistical way. This is also justified by the fact that in a hot plasma, the number of lines is huge, and therefore the number of Zeeman transitions, arising from the splitting of spectral lines, is even greater, which makes the coalescence of the spectral features more important. Due to the other physical broadening mechanisms, the Zeeman components can not be resolved (Doron et al., 2014).

In the presence of a magnetic field $B$, a level $\alpha J_1$ (energy $E_1$) splits into $2J_1+1$ states $M_1$ ($-J_1\leq M_1\leq J_1$) of energy $E_1+\mu_Bg_1M_1$ , $\mu_B$ being the Bohr magneton and $g_1$ the Land\'e factor in intermediate coupling (provided by RCG routine). Each line splits in three components associated to selection rule $\Delta M$=$q$, where $q$=0 for a $\pi$ component and $\pm 1$ for a $\sigma_{\pm}$ component. The intensity of a component can be characterized by the strength-weighted moments of the energy distribution. The $n^{th}-$ order moment reads

\begin{equation}
\mu_n\left[q\right]=3\sum_{M_1,M_2}\threejm{J_1}{1}{J_2}{-M_1}{-q}{M_2}^2\left(E_2-E_1+\mu_BB\left[g_2M_2-g_1M_1\right]\right)^n,
\end{equation}

\noindent which can be evaluated analytically (Pain \& Gilleron, 2012a \& 2012b), using graphical representation of Racah algebra or Bernoulli polynomials (Mathys \& Stenflo, 1987). 

\begin{figure}[ht]
\vspace{17mm}
\begin{center}
\includegraphics[width=10cm]{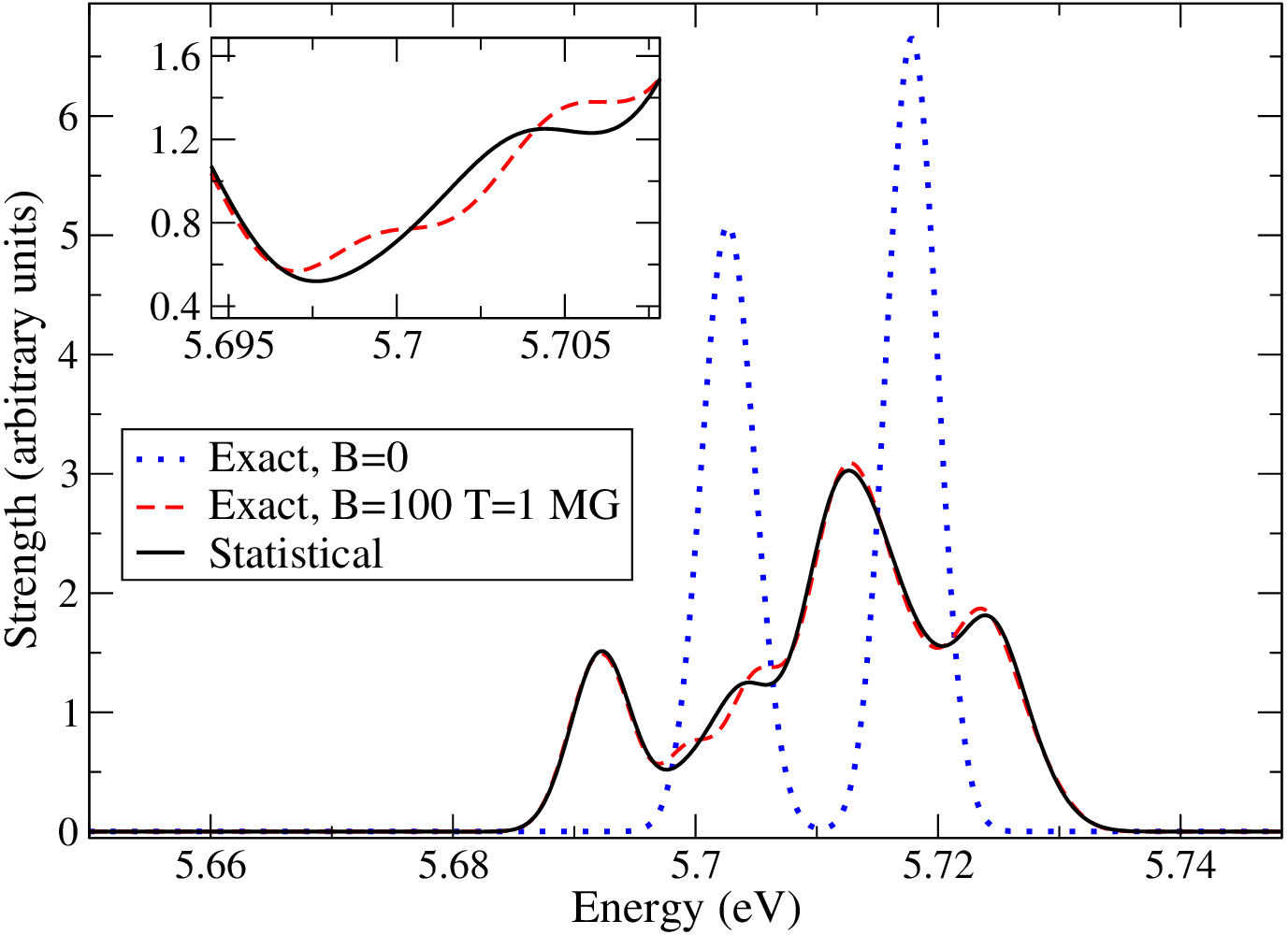}
\end{center}
\vspace{7mm}
\caption{Effect of a 1 MG magnetic field on triplet transition $1s2s~^3S\rightarrow 1s2p~^3P$ of carbon ion C$^{4+}$ with a convolution width (FWHM) of 0.005 eV. The observation angle $\theta$ is such that $\cos^2(\theta)=1/3$.\label{C}}
\label{fig1}
\end{figure}

\begin{figure}
\vspace{7mm}
\begin{center}
\includegraphics[width=10cm]{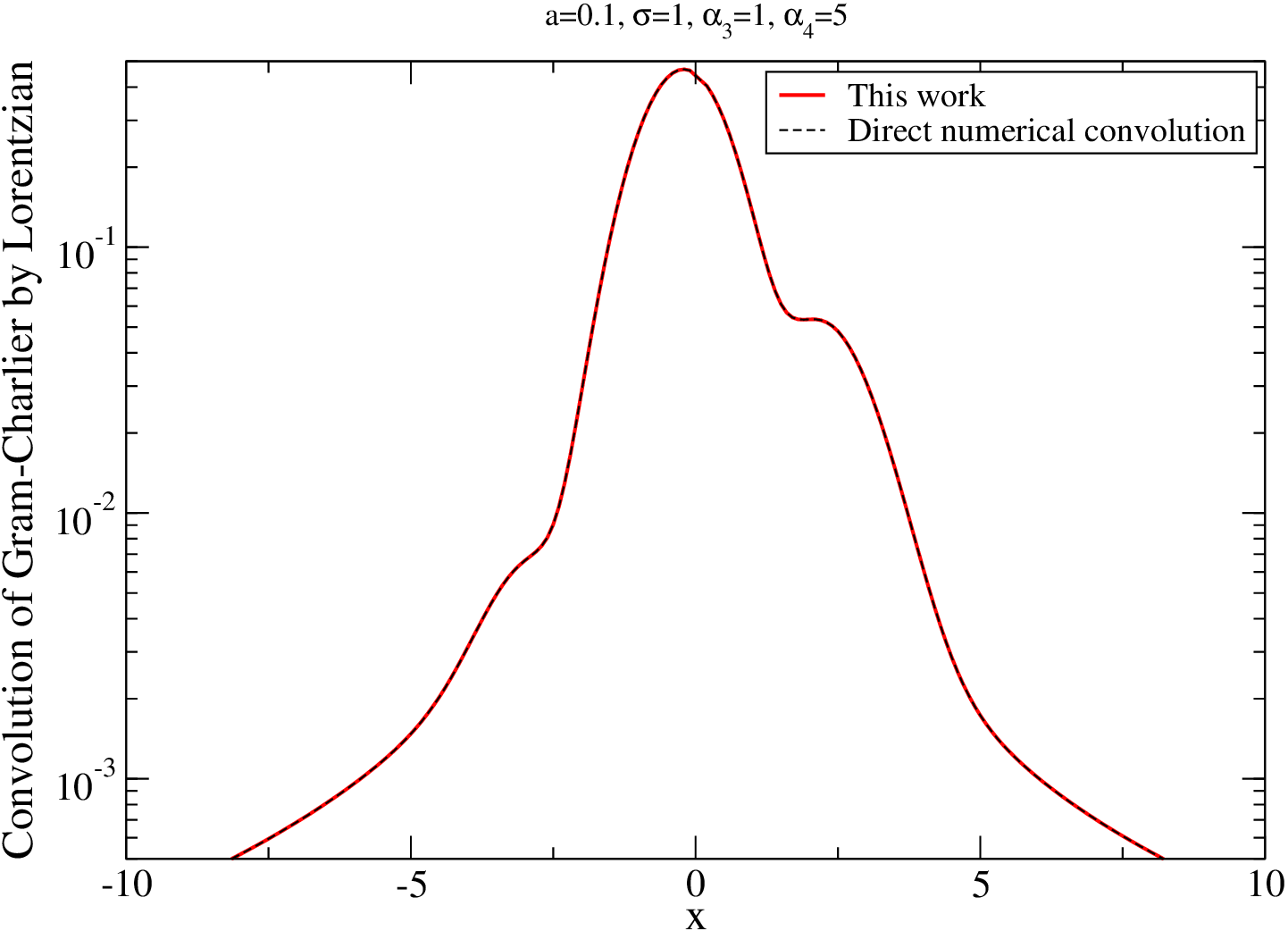}
\end{center}
\vspace{7mm}
\caption{(Color online) Convolution of Gram-Charlier expansion series with a Lorentzian. The parameters are $a$=0.1, $\sigma$=1, $\alpha_3$=1 and $\alpha_4$=5. The present approach (red curve) relying on a cubic-spline representation of the Gaussian and the direct numerical convolution (dashed black curve) are superimposed.\label{gcv_log}}
\end{figure}

\begin{table*}[t]
\begin{center}
\begin{tabular}{|c|c|c|c|}\hline
\multicolumn{2}{|c|}{} & $J_2=J_1$ & $J_2=J_1\pm 1$ \\\hline
$\sigma_q$ & $\alpha_3$ &  \multicolumn{2}{c|}{$(-1)^qq\left(J_1-J_2\right)\mathrm{sgn}\left[g_1-g_2\right]\frac{2\sqrt{5}}{3\sqrt{3}}\frac{J_>}{\sqrt{J_<\left(J_>+1\right)}}$} \\\cline{2-4}
& $\alpha_4$ & $\frac{5}{7}\left(\frac{12J_1\left(J_1+1\right)-17}{4J_1\left(J_1+1\right)-3}\right)$ & $\frac{5}{21}\left(13-\frac{4}{J_<\left(J_>+1\right)}\right)$ \\ 
\hline
$\pi$ & $\alpha_3$ & \multicolumn{2}{c|}{0} \\\cline{2-4}
& $\alpha_4$ & $\frac{25}{7}\left(\frac{3\left[\left(J_1+2\right)J_1^2-1\right]J_1+1}{\left[1-3J_1\left(J_1+1\right)\right]^2}\right)$ & $\frac{5}{7}\left(3-\frac{2}{J_<\left(J_>+1\right)}\right)$ \\\hline
\end{tabular}
\caption{Values of $\alpha_3$ and $\alpha_4$ of the Zeeman components. $J_<=\min(J_1,J_2)$ and $J_>=\max(J_1,J_2)$. $\mathrm{sgn}\left[x\right]$ is the sign of $x$.}
\label{tab:a}
\end{center}
\end{table*}

\begin{table*}[t]
\begin{center}
\begin{tabular}{|c|c|}\hline
$a_k$ & $e^{-(k+1)^2/2}\left[k^2(k+2)^2+e^{k+1/2}\left(k^2-1\right)^2\right]$ \\ \hline
$b_k$ & $-e^{-(k+1)^2/2}k(k+1)\left[8+3k+e^{k+1/2}(3k-5)\right]$ \\\hline
$c_k$ & $e^{-(k+1)^2/2}\left[4+10k+3k^2+e^{k+1/2}\left(3k^2-4k-3\right)\right]$ \\\hline
$d_k$ & $-e^{-(k+1)^2/2}\left[3+e^{k+1/2}(k-2)+k\right]$ \\\hline
\end{tabular}
\caption{Expression of the coefficients $a_k$, $b_k$, $c_k$ and $d_k$ involved in the cubic-spline representation of the Gaussian (Eq. (\ref{cubic})) for $k\le u\le k+1$.}\label{tab:c}
\end{center}
\end{table*}

\begin{table*}[t]
\begin{center}
\begin{tabular}{|c|c|}\hline
$i$ & $\gamma_{p,i}$ \\ \hline \hline 
0 & $\left[1+\frac{\alpha_4-3}{8}\right]a_p$ \\\hline
1 & $-\frac{\alpha_3}{2}a_p+\left[1+\frac{\left(\alpha_4-3\right)}{8}\right]b_p$ \\\hline
2 & $-\frac{\left(\alpha_4-3\right)}{4}a_p-\frac{\alpha_3}{2}b_p+\left[1+\frac{\left(\alpha_4-3\right)}{8}\right]c_p$ \\\hline
3 & $\frac{\alpha_3}{6}a_p-\frac{\left(\alpha_4-3\right)}{4}b_p-\frac{\alpha_3}{2}c_p+\left[1+\frac{\left(\alpha_4-3\right)}{8}\right]d_p$ \\\hline
4 & $\frac{\left(\alpha_4-3\right)}{24}a_p+\frac{\alpha_3}{6}b_p-\frac{\left(\alpha_4-3\right)}{4}c_p-\frac{\alpha_3}{2}d_p$ \\\hline
5 & $\frac{\left(\alpha_4-3\right)}{24}b_p+\frac{\alpha_3}{6}c_p-\frac{\left(\alpha_4-3\right)}{4}d_p$ \\\hline
6 & $\frac{\left(\alpha_4-3\right)}{24}c_p+\frac{\alpha_3}{6}d_p$ \\\hline
7 & $\frac{\left(\alpha_4-3\right)}{24}d_p$ \\\hline
\end{tabular}
\caption{Coefficients $\gamma_{p,i}$ involved in Eq. (\ref{gam}).}\label{tab:b}
\end{center}
\end{table*}

\subsection{Gram-Charlier distribution}

Gram-Charlier expansions are useful to model densities which are deviations from the normal one. The expansion is named after the Danish mathematician Jorgen P. Gram (1850-1916) and the Swedish astronomer Carl V. L. Charlier (1862-1934). Historical accounts of the origin of the Gram-Charlier expansion are given in Hald (2000) and Davies (2005). This expansion, that finds applications in many areas including finance (Jondeau \& Rockinger, 2001), analytical chemistry (Di Marco \& Bombi, 2001), spectroscopy (O'Brien, 1992) and astrophysics and cosmology (Blinnikov, 1998) reads

\begin{equation}
GC(u)=\frac{1}{\sqrt{2\pi v}}e^{-u^2/2}\left[\sum_{k=0}^{\infty}c_k\he_k\left(\frac{u}{\sqrt{2}}\right)2^{-k/2}\right],
\end{equation}

\noindent where $u=\left(h\nu-\mu_1\right)/\sqrt{v}$, $v=\mu_2-\left(\mu_1\right)^2$ being the variance. The polynomials $\he_k(x)$ can be expressed as

\begin{equation}
\he_k(x)=\frac{1}{2^{k/2}}\h_k\left(\frac{x}{\sqrt{2}}\right),
\end{equation}

\noindent where $\h_k(x)$ are the usual Hermite polynomials obeying the recurrence relation (Szego, 1939):

\begin{equation}
\h_{k+1}(x)=2x\h_k(x)-2k\h_{k-1}(x)
\end{equation}

\noindent initialized with $\h_0(x)$=1 and $\h_1(x)=2x$. The coefficients $c_k$ are given by

\begin{equation}
c_k=\sum_{j=0}^{\left[k/2\right]}\frac{(-1)^j}{j!(k-2j)!2^j}\alpha_{k-2j}
\end{equation}

\noindent where $[.]$ denotes the integer part and $\alpha_k$ is the dimensionless centered $k-$order moment of the distribution 
 
\begin{equation}
\alpha_k=\left(\sum_{p=0}^k\bin{k}{p}\mu_p\left(-\mu_1\right)^{k-p}\right)/v^{k/2}.
\end{equation}

\noindent A good representation of the Zeeman profile is obtained using, for each component, the fourth-order Gram-Charlier expansion series:

\begin{equation}\label{gc4}
\Psi_Z(u)=\frac{1}{\sqrt{2\pi v}}\exp\left(-\frac{u^2}{2}\right)\left[1-\frac{\alpha_3}{2}\left(u-\frac{u^3}{3}\right)+\frac{\left(\alpha_4-3\right)}{24}\left(3-6u^2+u^4\right)\right],
\end{equation}

\noindent where $\alpha_3$ (skewness) and $\alpha_4$ (kurtosis) quantify respectively the asymmetry and the sharpness of the component (see Table \ref{tab:a}) (Kendall \& Stuart, 1969). This approximate method was shown to provide quite a good description (see Fig. \ref{C}) of the effect of a strong magnetic field on spectral lines (Pain \& Gilleron, 2012a \& 2012b). The contribution of a magnetic field to an UTA can be taken into account roughly by adding a contribution $2/3\;(\mu_BB)^2\approx 3.35\; 10^{-5}$ [$B$(MG)]$^2$ eV$^2$ to the statistical variance. 

When all the other broadening mechanisms (statistical, Doppler, ionic Stark) are described by a Gaussian, the resulting profile (convolution of a Gaussian by Gram-Charlier) remains a Gram-Charlier function with modified moments. However, electron collisional broadening is usually modeled by a Lorentzian function

\begin{equation}
L(h\nu,a)=\frac{a}{\pi}\frac{1}{a^2+\left(h\nu\right)^2},
\end{equation}

\noindent as well as natural width. The convolution of a Gaussian by a Lorentzian leads to a Voigt profile (Voigt, 1912, Matveev, 1972 \& 1981, Ida et al., 2000) but in the presence of a magnetic field, the problem is more complicated, since the numerical cost of the direct numerical convolution of a Lorentzian with Gram-Charlier function is prohibitive, due to the huge number of lines involved in the computation. It reads

\begin{equation}\label{conv}
C(t)=\left(GC\otimes L\right)(t)=\int_{-\infty}^{\infty}GC\left(u\right)L(t-u,\lambda)du
\end{equation}

\noindent where $t=h\nu/\sigma,$ $\sigma=\sqrt{v}$ being the standard deviation of the distribution and $\lambda=a/\sigma$.

\subsection{Convolution of a Lorentzian function with Gram-Charlier expansion series}

The convolution product (\ref{conv}) requires the evaluation of a cumbersome integral which reads

\begin{equation}\label{dep}
C(t)=\frac{1}{\pi\sqrt{2\pi}}\frac{\lambda}{\sigma}\int_{-\infty}^{\infty}\frac{e^{-u^2/2}}{\lambda^2+(t-u)^2}\times\left[\sum_{k=0}^{\infty}c_k\he_k\left(\frac{u}{\sqrt{2}}\right)2^{-k/2}\right]du.
\end{equation}

\noindent In order to fasten the calculation, the Gaussian is sampled at the points $u=-m,-m+1,\cdots,0,\cdots,m-1,m$ (in practice we use $m$=6) and interpolated using cubic splines (de Boor, 1978) on each interval $\left[k,k+1\right]$ by the formula

\begin{equation}\label{cubic}
e^{-u^2/2}=a_k+b_k~u+c_k~u^2+d_k~u^3.
\end{equation}

\noindent The coefficients $a_k$, $b_k$, $c_k$ and $d_k$ in the interval $\left[k,k+1\right]$ are determined by the continuity of the function and its derivative at the points $u=k$ and $u=k+1$. The resulting expressions are given in Table \ref{tab:c}. The Gaussian is assumed to be zero for $|u|>m$. Limiting Gram-Charlier expansion series to fourth order (see Eq. (\ref{gc4})), one now has to deal with the convolution of a Lorentzian by a polynomial of order 7. This can be written\footnote{In the general case, the upper bound of the sum over $k$ is equal to $N$+3, where $N$ is the order of the Gram-Charlier expansion series.}

\begin{equation}\label{gam}
C(t)=\frac{1}{\pi\sqrt{2\pi}\sigma}\sum_{p=-m}^{m-1}\sum_{k=0}^7\gamma_{p,k}~S_{p,k}(t),
\end{equation}

\noindent where the coefficients $\gamma_{p,k}$ of the polynomial are given in Table \ref{tab:b}, and

\begin{eqnarray}
S_{p,k}(t)&=&\int_{p}^{p+1}\frac{u^k}{\lambda^2+\left(t-u\right)^2}du\nonumber\\
&=&\frac{1}{\lambda}\sum_{\ell=0}^k\bin{k}{\ell}\lambda^{\ell}t^{k-\ell}\left[\phi_{\ell}\left(\tfrac{p+1-t}{\lambda}\right)-\phi_{\ell}\left(\tfrac{p-t}{\lambda}\right)\right].
\end{eqnarray}

\noindent The function $\phi_{\ell}(w)$ is equal to

\begin{equation}
\phi_{\ell}(w)=\frac{w^{\ell+1}}{\ell+1}~_2F_1
\left(\begin{array}{l}
1,\frac{\ell+1}{2}\\
\frac{\ell+3}{2}
\end{array};-w^2\right),
\end{equation}

\noindent where $_2F_1$ is a hypergeometric function, but can be efficiently obtained using the recurrence relation

\begin{equation}
\phi_{\ell}(w)=w^{\ell-2}-\phi_{\ell-2}(w)
\end{equation}

\noindent with

\begin{equation}
\phi_0(w)=\arctan(w)\;\;\;\mathrm{and}\;\;\;\phi_1(w)=\frac{1}{2}\ln\left[1+w^2\right].
\end{equation}

\noindent Such a method provides fast and accurate results, even for very asymmetrical and sharp Gram-Charlier distribution (see Fig. \ref{gcv_log}). The total line profile results from the convolution of $\Psi_Z$ with other broadening mechanisms. If $\sigma\leq a/10$, we take only the Lorentzian $L(h\nu,a)$. On the other hand, if $a\leq \sigma/150$, we keep the Gaussian. If one has to convolve $C(t)$ by  an additional Gaussian of variance $\sigma'$ (representing Doppler broadening for instance), $\sigma$, $\alpha_3$ and $\alpha_4$ must be replaced respectively by:

\begin{equation}
\left\{\begin{array}{l}
\tilde{\sigma}=\sqrt{\sigma^2+\sigma'^2}\\
\tilde{\alpha}_3=\alpha_3\left(\frac{\tilde{\sigma}}{\sigma}\right)^3\\
\tilde{\alpha}_4=\alpha_4\left(\frac{\tilde{\sigma}}{\sigma}\right)^4
\end{array}\right.
\end{equation}

\begin{figure}
\vspace{7mm}
\begin{center}
\includegraphics[width=10cm]{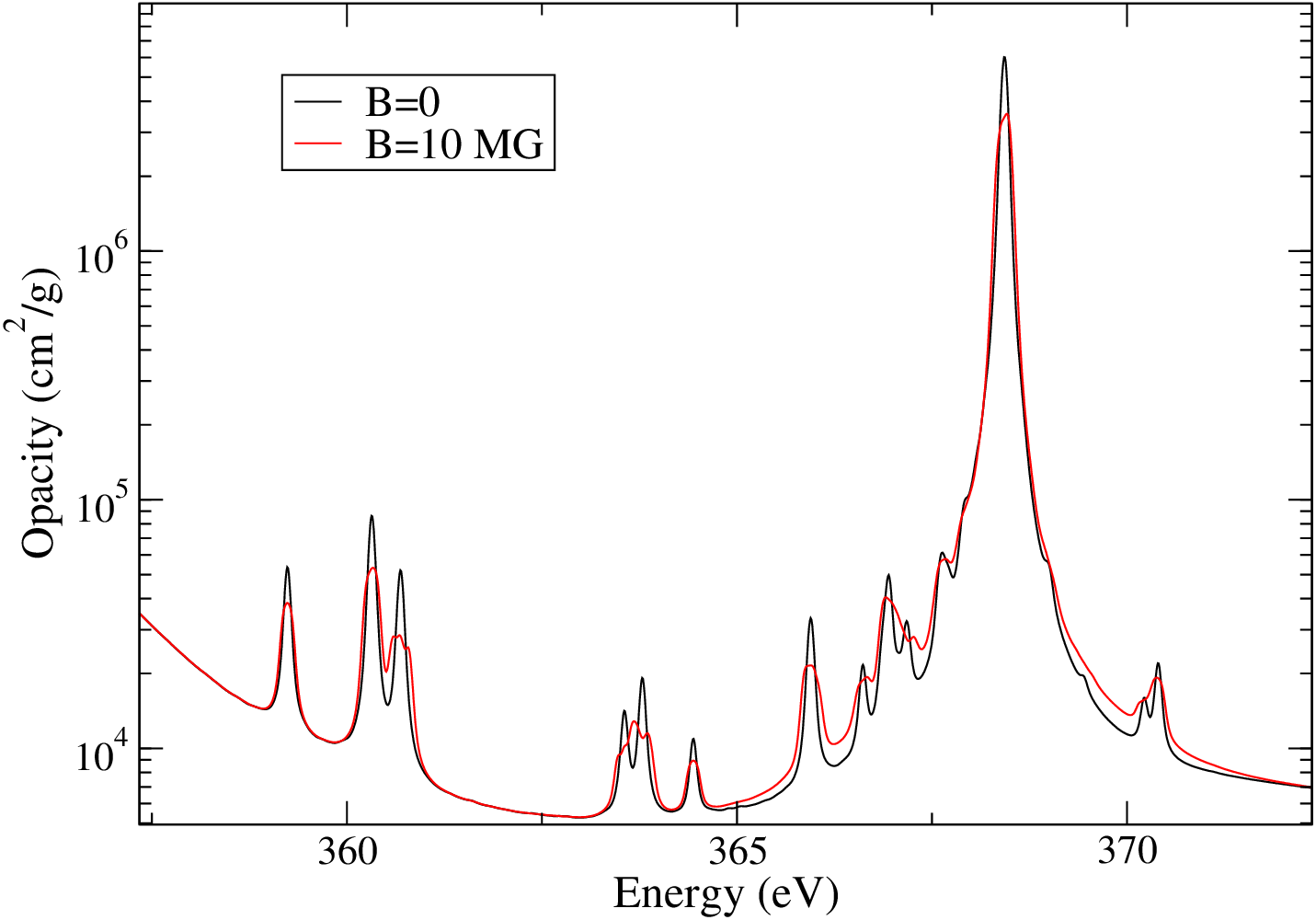}
\end{center}
\vspace{7mm}
\caption{(Color online) SCO-RCG calculations (transitions $1s\rightarrow 2p$) with and without magnetic field for a carbon plasma at $T$=50 eV and $\rho$=10$^{-2}$ g.cm$^{-3}$ (conditions typical to ICF).\label{fig1_FCI}}
\end{figure}

\begin{figure}
\vspace{7mm}
\begin{center}
\includegraphics[width=10cm]{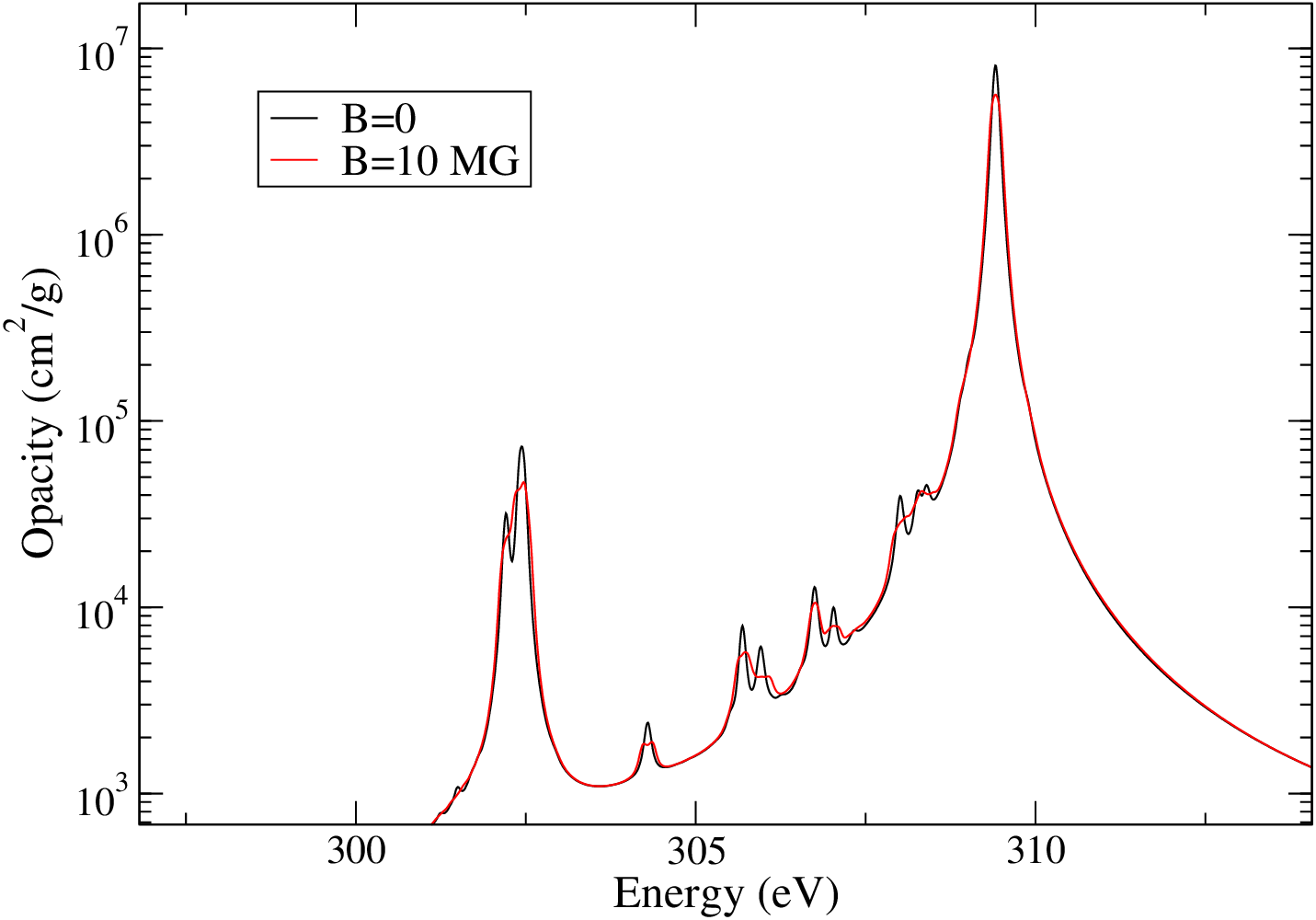}
\end{center}
\vspace{7mm}
\caption{(Color online) SCO-RCG calculations (transitions $1s\rightarrow 2p$) with and without magnetic field for a carbon plasma at $T$= 50 eV and $\rho$=10$^{-2}$ g.cm$^{-3}$ (condidions typical to ICF) in a spectral range close to the one of Fig. \ref{fig1_FCI}.\label{fig2_FCI}}
\end{figure}

\noindent Figs. \ref{fig1_FCI} and \ref{fig2_FCI} illustrate the impact of a 10 MG magnetic field (typical of ICF) in the XUV range for a carbon plasma at $T$=50 eV and $\rho$=0.01 g/cm$^3$.   

\section{Conclusion}

By combining different degrees of approximation of the atomic structure (levels, configurations and superconfigurations), the SCO-RCG code allows us to explore a wide range of applications, such as the calculation of Rosseland means, the generation of opacity tables, or the spectroscopic interpretation of high-resolution spectra. The Partially Resolved Transition Array model was recently adapted to the hybrid statistical / detailed approach in order to reduce the statistical part and speed up the calculations. An approximate approach providing a fast and quite accurate estimate of the effect of an intense magnetic field on opacity was also implemented. The formalism requires the moments of the Zeeman components of a line $\alpha J\rightarrow\alpha'J'$, which can be obtained analytically in terms of the quantum numbers and Land\'{e} factors and the profile is modeled by the fourth-order A-type Gram-Charlier expansion series. We also proposed a fast and accurate method to perform the convolution of this Gram-Charlier series with a Lorentzian function. Such an algorithm is useful in order to account for distorsions of the Voigt profile, since the direct numerical evaluation of the integral becomes rapidly prohibitive. More generally, It can be helpful for models relying on the theory of moments (Bancewicz \& Karwowski, 1987), used in most opacity and emissivity codes. In the future, we plan to extend the statistical modeling of Zeeman effect using temperature-dependent moments (see Appendix) and to improve the treatment of Stark broadening in order to increase the capability of the code as concerns K-shell spectroscopy. 

\clearpage

\appendix

\section{Temperature-dependent moments of Zeeman Hamiltonian}

Polarized synchrotron radiation can be used to determine the magnitude, the orientation and the temperature and magnetic-field dependence of the local rare-earth magnetic moment in magnetically ordered materials. Thole et al. (1985) proposed a theory which predicts an anomalously large magnetic dichro\"ism in the $M_{4,5}$ X-ray absorption-edge structure. The square of the matrix element of an optical dipole transition from a state $\alpha J M$ to a final state $\alpha'J'M'$ is, according to the Wigner-Eckart theorem, proportional to the square of the 3j symbol times the reduced matrix element (line strength in the absence of a magnetic field):

\begin{equation}
S_{\alpha J M,\alpha'J'M'}=\threejm{J}{1}{J'}{M}{q}{-M'}^2|\langle\alpha J||C^{(1)}||\alpha'J'\rangle|^2,
\end{equation}

\noindent where $\alpha$ labels different levels of equal $J$ and $q$=0 for light polarized in the field direction, $q=\pm 1$ for right- or left-circularly polarized light perpendicular to the field direction. The partition function associated to a particular level $\alpha J$ reads

\begin{equation}
Z_{\alpha J}=\sum_{M=-J}^Je^{-C_{\alpha J}M}=\frac{\sinh\left[C_{\alpha J}(J+1/2)\right]}{\sinh(C_{\alpha J}/2)},
\end{equation}

\noindent where $C_{\alpha J}=\mu_B Bg_{\alpha J}/(k_BT)$, $\mu_B$ being the Bohr magneton, $B$ the intensity of the magnetic field and $g_{\alpha J}$ the Land\'e factor of level $\alpha J$ in intermediate coupling. The first order moment can be expressed as

\begin{eqnarray}
\langle M\rangle&=&\frac{1}{Z_{\alpha J}}\sum_{M=-J}^JMe^{-C_{\alpha J}M}=-(J+1/2)\coth\left[C_{\alpha J}(J+1/2)\right]+\frac{1}{2}\coth(C_{\alpha J}/2)\nonumber\\
&=&-J~B_J(C_{\alpha J}J),
\end{eqnarray}

\noindent where $B_J$ denotes Brillouin's function (Darby, 1967, Subramanian, 1986):

\begin{equation}
B_J(x)=\frac{2J+1}{2J}\coth\left(\frac{2J+1}{2J}x\right)-\frac{1}{2J}\coth\left(\frac{x}{2J}\right).
\end{equation}

\noindent For $\langle M^2\rangle$, one has 

\begin{equation}
\langle M^2\rangle=\frac{1}{Z_{\alpha J}}\sum_{M=-J}^JM^2e^{-C_{\alpha J}M}=J(J+1)+\langle M\rangle\coth(C_{\alpha J}/2)
\end{equation}

\noindent and the higher-order moments can be obtained using the following relation:

\begin{equation}
\langle M^n\rangle=\frac{(-1)^n}{Z_{\alpha J}}\frac{\partial^n Z_{\alpha J}}{\partial C_{\alpha J}^n}=\langle M\rangle\langle M^{n-1}\rangle-\frac{\partial}{\partial C_{\alpha J}}\langle M^{n-1}\rangle,
\end{equation}

\noindent where $n\in\mathbb{N}$.

\end{document}